\documentclass[twocolumn,showpacs,preprintnumbers,amsmath,amssymb]{revtex4}

\usepackage{graphicx}
\usepackage{dcolumn}
\usepackage{bm}

\begin{document}


\title{
Distribution Function of Electron Velocity 
Perpendicular to the Driving Force 
in a Uniform Nonequilibrium Steady State
}

\author{Tatsuro Yuge}
\email{yuge@m.tains.tohoku.ac.jp}
\affiliation{
IIAIR, Tohoku University,  
Aoba-ku, Sendai, Miyagi 980-8578, Japan
}

\date{\today}

\begin{abstract}
A macroscopically uniform model of a two-dimensional electron system 
is proposed to study nonequilibrium properties of electrical conduction.
By molecular dynamics simulation, the steady state distribution function $P_y$ 
of electron velocity in a direction perpendicular 
to an external driving force is calculated.
An explicit form of $P_y$ is determined 
within the accuracy of the numerical simulation, 
which fits the numerical data well even in the regime 
where a local equilibrium description is not valid.
Although the entire structure of $P_y$ is different from 
that of a local  equilibrium distribution function, 
the asymptotic structure of the tails of $P_y$ in the limit of large absolute values 
of the velocity is identical to that of a Maxwell distribution function 
with a temperature which is different from that in the equilibrium state 
and the kinetic temperature in the steady state. 
\end{abstract}



\maketitle

One of the central roles of statistical mechanics is to provide 
a principle for calculating the probability distribution $\rho$ 
of a microscopic state without solving equations of motion.
In equilibrium states it is given as the canonical distribution. 
In nonequilibrium steady states (NESSs), by contrast, 
such a principle is not known.

In studies of nonequilibrium systems with some concrete models, 
it is generally easier to compute a single-particle distribution function, 
such as a velocity distribution function (VDF), 
than to calculate $\rho$ itself.
In equilibrium states, the VDF is, of course, the Maxwell distribution.
Although the VDF contains less information than $\rho$ 
since the VDF is a marginal probability distribution, 
it yields a restriction on $\rho$ 
and knowledge of the VDF is thus expected to provide a clue 
to finding a principle of nonequilibrium statistical mechanics.

VDFs have been investigated mainly in models described by kinetic equations, 
such as the Boltzmann equation and the Bhatnagar-Gross-Krook equation 
\cite{KimHayakawa}.
Such kinetic models are said to be valid for dilute gas systems.
Furthermore, VDFs in most studies on those models 
are given as a power series of degrees of nonequilibrium 
(such as a temperature gradient).
From these facts, 
it may be interesting to investigate a model other than kinetic models 
using a method without power series expansions.

One of the approaches in such studies is 
molecular dynamics (MD) simulation on models 
which obey microscopic equations of motion.
VDFs have been investigated 
with MD simulations in shear-flowing systems \cite{LooseHess} 
and thermal-conducting lattice systems \cite{AokiKusnezov,UedaTakesue}.
In those studies, deviations from local equilibrium distribution functions 
(LEDFs) have mainly been considered.
However, sufficient characterization of VDFs 
from the viewpoint of statistical mechanics 
has not yet been achieved.
Another example of MD simulations 
was a study of the distribution function of energy currents 
carried by a single particle in thermal-conducting particle systems 
\cite{Yukawa_et_al}.
In that work, the authors found that the energy current distribution function 
has tails at large absolute values of the current, 
which are  well described by equilibrium distributions 
with appropriate temperatures.
This tail structure might be a universal property 
of steady state distribution functions.
That is, for a wide range of nonequilibrium systems, 
including thermal-conducting and other systems, 
tails of distributions such as VDFs and energy current distributions 
might asymptotically approach certain equilibrium distributions, 
although the complete structures would be different 
from those of the equilibrium distributions.

In the present paper, to demonstrate the above idea, 
we investigate a VDF in an electrically conducting system.
We introduce a model 
in which a macroscopically uniform NESS is realized, 
and perform MD simulation to calculate a VDF in a direction 
perpendicular to an external driving force.
We find an explicit form of the VDF within simulation accuracy.
Introduction of a velocity-dependent ``reciprocal temperature'' 
helps to determine the form.

\begin{figure}[tb]
\begin{center}
\includegraphics[width=.85\linewidth]{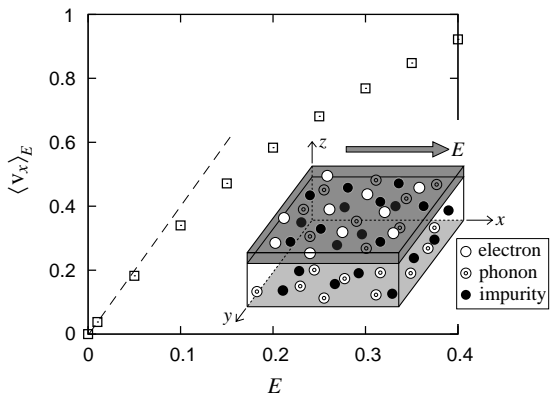}
\end{center}
\caption{
Average of electron velocity in the $x$-direction, plotted against $E$.
The dashed line represents the slope of the linear response.
Inset: schematic diagram of the simulation system.
Electrons are confined in a 2D plane (dark gray).
Under the 2D electron system there is a three-dimensional phonon system, 
the bottom of which is a thermal wall (light gray).
Impurities are distributed throughout all regions of the system.
An external electric field $E$ is applied in the $x$-direction.
}
\label{model}
\end{figure}

\textit{Model.}
The system of the simulation is composed of electrons, 
phonons, and impurities, which are modeled as classical particles.
A schematic diagram of the system is shown in the inset of Fig.~\ref{model}.
The linear dimension in the $\alpha$-direction of the system 
is denoted by $L_{\alpha}$ ($\alpha = x,y,z$).
Electrons (whose mass, charge, and total number are denoted by 
$m_e$, $e$, and $N_e$) 
can move only in a two-dimensional (2D) $x$-$y$ plane (shown in dark gray) 
which is located on the top ($z=L_z$) of the three-dimensional (3D) system.
Phonons (whose mass and total number are denoted by $m_p$ and $N_p$)
can move throughout the entire 3D system.
Impurities (whose total number is denoted by $N_i$) 
are fixed at random positions, 
and play the role of a random potential.
$N_i^{\rm 2D}$ of the impurities are uniformly distributed 
throughout the 2D electron system 
and $N_i^{\rm 3D}$ throughout the 3D system under the 2D system
($N_i^{\rm 2D} + N_i^{\rm 3D} = N_i$).
Periodic boundary conditions are imposed in the $x$- and $y$-directions.
The boundary conditions in the $z$-direction, applied only to phonons, 
are an elastic potential wall for the top boundary ($z=L_z$) 
and a thermal wall (shown in light gray) with temperature $T_0$ 
for the bottom boundary ($z=0$).
We apply an external electric field $E$ (acting on electrons) 
uniformly in the $x$-direction 
to drive the system to a nonequilibrium state.
We assume that interactions among all kinds of particles are present.
The interaction potential between the $k$-th and $l$-th particles 
is given by $\phi_{kl} = \mathcal{Y} (\max \{ 0, d_{kl} \})^{5/2}$.
Here, $\mathcal{Y}$ is a constant of interaction strength, 
and $d_{kl}=R_k +R_l - r_{kl}$ 
is the overlap of the potential ranges.
$R_k$ is the radius of the potential range 
($R_e$, $R_p$, and $R_i$ 
for an electron, phonon, and impurity, respectively), 
and $r_{kl}$ is the inter-center distance between the particles.
The energy supplied from $E$ to electrons is transferred to phonons 
by electron-phonon interactions and dissipates through the thermal walls, 
by which the system retains its energy balance.
In the simulations we take $e$, $R_e$, $m_e$, a reference energy, 
and the Boltzmann constant as the units, 
and fix the following parameters: $L_x=L_y=200$, $L_z=10$, 
$R_p=R_i=1$, $m_p=1$, $T_0=1$, $N_e=1000$, $N_p=4000$, 
$N_i^{\rm 2D}=500$, $N_i^{\rm 3D}=1000$, and $\mathcal{Y}=4000$.

This model is an extension of the two-dimensional model 
for the MD simulation of electrical conduction, 
which has been previously proposed in Ref.~\cite{YIS}.
A possible experimental situation corresponding to 
the model in Ref.~\cite{YIS} occurs when 
the edges of a 2D electron system are in contact with large sample holders 
(which have large heat capacity and good heat conduction).
On the other hand, 
the present model is a reflection of a typical experimental setup 
of a quasi-2D electron system in a semiconductor device.
That is, a 2D electron system is realized around the top of a bulk substrate 
(which is modeled by the 3D system of phonons and impurities) 
and the substrate is mounted on a large sample holder 
(which is modeled by the thermal wall).
The present model is macroscopically uniform 
in the $y$-direction as well as in the $x$-direction, 
whereas the previous one was uniform only in the direction parallel to $E$.
This enables us to investigate bulk properties 
in both the $x$- and $y$-directions.
It should also be noted that 
this model is isotropic in the $y$-direction 
and the net currents of particles and energy 
in this direction are absent on average.

In Fig.~\ref{model} we show $E$-dependence of 
the average electron velocity $\langle v_x \rangle_E$ 
in the $x$-direction.
Here $\langle \cdots \rangle_E$ denotes the average 
in the steady state for $E$.
A linear response is observed in the regime of small $E$ 
and a nonlinear response in the regime of large $E$.

\begin{figure}[tb]
\begin{center}
\includegraphics[width=.85\linewidth]{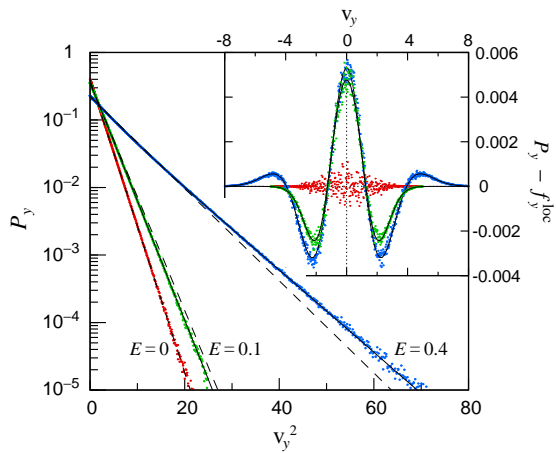}
\end{center}
\caption{(Color online) 
A semi-logarithmic plot of the distribution $P_y$, 
plotted against $v_y{}^2$.
The results for $E=0$, $0.1$, and $0.4$ are shown.
The data for $v_y>0$ and $v_y<0$ are overlapped.
The dashed lines represent the LEDFs $f_y^{\rm loc}$.
The solid lines depict fitting curves of Eq.~(\ref{P_y_fun}) 
for $E=0.1$ and $0.4$.
Inset: deviation $P_y(v_y) - f_y^{\rm loc}(v_y)$ 
from the LEDF.
Also shown are the fitting curves of Eq.~(\ref{P_y_fun}), 
with $f_y^{\rm loc}$ subtracted, 
for $E=0.1$ and $0.4$ (solid curves).
}
\label{Py}
\end{figure}

\textit{Deviation from LEDF.}
Because of the macroscopic spatial uniformity of the model, 
steady states in the model are also uniform.
That is, local quantities such as local number densities 
and local kinetic temperatures in steady states 
are almost independent of the position in the 2D electron plane.
We may therefore use electrons in the entire region of 
the plane to calculate a VDF of electrons.
In this paper we investigate distribution functions $P_y(v_y)$ 
of electron velocity $v_y$ in the $y$-direction.
[The corresponding results in the $x$-direction (parallel to $E$) 
will be presented elsewhere \cite{YS_dist_x};
these have more complicated structures and properties 
than those of $P_y$ presented here.]
In Fig.~\ref{Py} we show semi-logarithmic plots of 
$P_y$, plotted against $v_y{}^2$.
The data for positive $v_y$ 
almost completely overlap with those for negative $v_y$
in NESSs ($E=0.1$ and $0.4$), 
as well as in the equilibrium state ($E=0$).
This indicates that $P_y(v_y)$ is symmetric at $v_y = 0$, 
and is consistent with $\langle v_y \rangle_E = 0$, 
even in the nonlinear response regime.
These results are due to the isotropy in the $y$-direction 
of the model.

As an LEDF in the $y$-direction, we employ 
a naive distribution function defined by 
\begin{equation}
f_y^{\rm loc} (v_y) = \sqrt{m_e / 2\pi T_e^y} 
\exp \left( -m_e v_y^2 / 2T_e^y \right), 
\label{local_eq_y}
\end{equation}
where $T_e^y = m_e \langle v_y{}^2 \rangle_E$ is 
a kinetic temperature in this direction 
(this is simply a Maxwell distribution with temperature $T_e^y$).
In Fig.~\ref{Py} we also plot $f_y^{\rm loc}$ 
(in this plot zero-mean Gaussian distributions are drawn as straight lines).
We observe that the tails of $P_y$ and $f_y^{\rm loc}$ 
are different for large values of $E$, 
whereas they are almost the same for $E=0$.
Moreover in the inset of  Fig.~\ref{Py}, 
we plot the deviation $P_y(v_y) - f_y^{\rm loc}(v_y)$.
From this figure we see that the difference around $v_y = 0$ 
is also large in the NESSs.
The symmetric behavior of the deviation is again consistent 
with the isotropic nature of the model.

\begin{figure}[tb]
\begin{center}
\includegraphics[width=.85\linewidth]{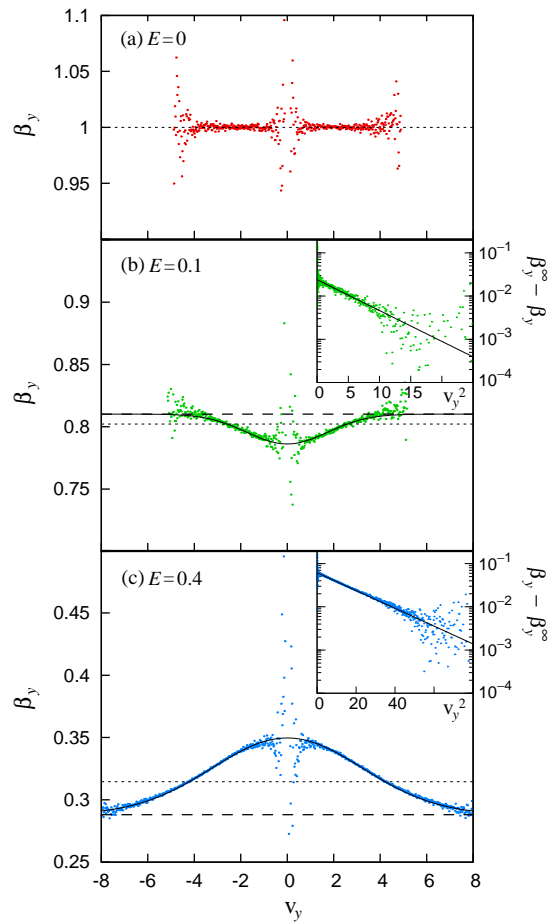}
\end{center}
\caption{(Color online) 
Mains: $\beta_y$ for (a) $E=0$, (b) $E=0.1$, and (c) $E=0.4$, 
plotted against $v_y$.
The dotted and dashed lines represent 
$1/T_e^y$ and $\beta_y^{\infty}$, respectively 
($1/T_e^y \simeq \beta_y^{\infty} \simeq 1/T_0 = 1$ when $E=0$).
The solid lines are fitting curves of Eq.~(\ref{beta_fun}).
Insets: semi-logarithmic plots of (b) $\beta_y^{\infty} - \beta_y$ 
and (c) $\beta_y - \beta_y^{\infty}$, respectively, 
plotted against $v_y{}^2$.
The data for $v_y>0$ and $v_y<0$ are overlapped.
The solid straight lines represent fitting curves of Eq.~(\ref{beta_fun}) 
(subtracted from $\beta_y^{\infty}$ in (b) 
and reduced by $\beta_y^{\infty}$ in (c), respectively).
}
\label{beta}
\end{figure}

\textit{Explicit form of $P_y$.}
To explore the functional form of $P_y$ 
we next investigate a $v_y$-dependent ``reciprocal temperature'',
$\beta_y (v_y)$, which is defined by 
\begin{equation}
\beta_y (v_y) = -\frac{2\ln \left[ P_y(v_y)/P_y(0) \right] }{m_e v_y{}^2} .
\label{beta_def}
\end{equation}
If $P_y$ is a Maxwell distribution, 
$\beta_y$ is independent of $v_y$ 
and is equal to the reciprocal temperature of the Maxwell distribution.
In Fig.~\ref{beta} we show $v_y$-dependence of $\beta_y$.
Figure~\ref{beta} (a) is a result for the equilibrium state ($E=0$).
In this case $\beta_y$ is almost independent of $v_y$
and is nearly equal to the reciprocal temperature $1/T_0$ ($=1$) 
of the thermal wall within the precision of the numerical simulation.
This is consistent with the fact that in an equilibrium state, 
$P_y$ is a Maxwell distribution with $T_0$.
Although the data fluctuations are rather large 
for $v_y$'s close to zero and in the tails, 
they are simply due to the restricted accuracy 
of the numerical simulation in these regions 
and they can be expected to become smaller 
as we average over greater numbers of simulation samples.

Figures~\ref{beta} (b) and (c) show the results 
for NESSs ($E=0.1$ and $0.4$, respectively).
As $|v_y|$ becomes large, 
$\beta_y$ varies (increasing in (b) and decreasing in (c)) monotonically 
and tends to converge to a certain constant value $\beta_y^{\infty}$, 
which is different from both $1/T_0$ and $1/T_e^y$.
This implies that $P_y$ approaches a Maxwell distribution 
as $|v_y| \to \infty$.
Furthermore, we find that the overall behavior of $\beta_y$ 
is well described by a Gaussian function.
That is, $\beta_y$ is well fitted by 
\begin{equation}
\beta_y (v_y) = B \exp (-D v_y{}^2) + \beta_y^{\infty}, 
\label{beta_fun}
\end{equation}
where $\beta_y^{\infty}$, $B$, and $D$ are fitting parameters. 
In the insets of Figs.~\ref{beta} (b) and (c), to see this more clearly, 
we show semi-logarithmic plots of $|\beta_y - \beta_y^{\infty}|$ 
versus $v_y{}^2$.
We observe exponential decay behavior 
of $|\beta_y - \beta_y^{\infty}|$ as functions of $v_y{}^2$, 
which indicates that the variation of $|\beta_y - \beta_y^{\infty}|$ 
is almost equivalent to that of Gaussian functions of $v_y$.
Although the data for $v_y$'s close to zero and in the tails 
fluctuate rather violently, they would converge to a single curve 
as the number of samples increases.
From Eqs.~(\ref{beta_def}) and (\ref{beta_fun}) 
we thus obtain an explicit functional form of $P_y$ as 
\begin{equation}
P_y (v_y) = P_y(0) \exp \left[ - \left( \beta_y^{\infty} 
+ B e^{-D v_y{}^2} \right) \frac{m_e v_y{}^2}{2} \right] .
\label{P_y_fun}
\end{equation}
This is a main result of the present paper.
In Fig.~\ref{Py} we plot curves of Eq.~(\ref{P_y_fun}) 
for $E=0.1$ and $0.4$ as solid lines.
These also reveal that Eq.~(\ref{P_y_fun}) is a good 
fitting function of $P_y$ for almost all values of $v_y$.

\begin{figure}[tb]
\begin{center}
\includegraphics[width=.75\linewidth]{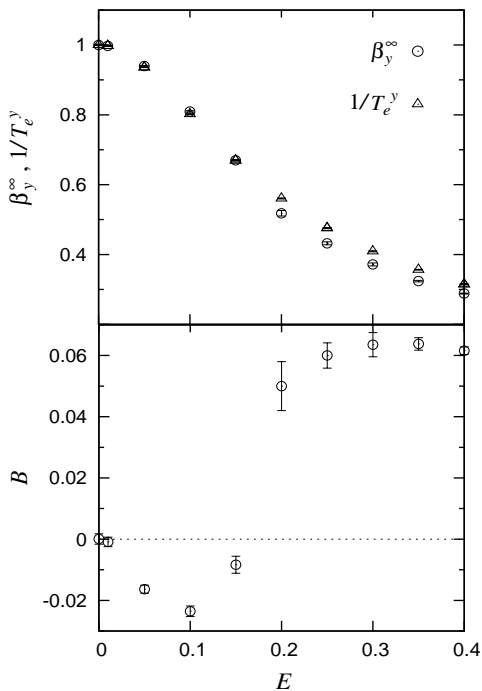}
\end{center}
\caption{
Top: $E$-dependence of $\beta_y^{\infty}$ (circles) 
and $1/T_e^y$ (triangles).
Bottom: $E$-dependence of $B$.
}
\label{parm}
\end{figure}

\textit{$E$-dependence.}
Figure~\ref{parm} shows the $E$-dependences 
of parameters in Eq.~(\ref{P_y_fun}).
In the top of Fig.~\ref{parm} we compare $\beta_y^{\infty}$ and $1/T_e^y$.
Although $E$-dependences of these two quantities are similar, 
their detailed values are different.
$\beta_y^{\infty} > 1/T_e^y$ at small values of $E$ 
(as can be seen also in Fig.~\ref{beta} (b)), 
whereas $\beta_y^{\infty} < 1/T_e^y$ at large values of $E$ 
(as is seen also in Fig.~\ref{beta} (c)).
In the bottom of Fig.~\ref{parm} we show the result for $B$.
There exists a value $E_c$ of the electric field 
such that $B < 0$ if $0 < E < E_c$ and $B > 0$ if $E > E_c$ 
($E_c \simeq 0.15$ in the parameter values of the present simulation).
This is consistent with the result for $\beta_y^{\infty}$, 
which is mentioned above. 
That is, $\beta_y^{\infty} > 1/T_e^y$ if $0 < E < E_c$ 
and $\beta_y^{\infty} < 1/T_e^y$ if $E > E_c$.

\textit{Summary and discussion.}
In this paper we have introduced a model 
of a two dimensional classical electron system 
which is macroscopically uniform.
In the model with an external driving force (electric field $E$), 
a spatially uniform nonequilibrium steady state (NESS) is realized.
By using molecular dynamics simulations 
we have investigated the distribution function $P_y(v_y)$ 
of an electron velocity $v_y$ in a direction perpendicular 
to $E$ in a steady state.
When the driving force is large, 
the local equilibrium (LE) description is not valid.
We have determined an explicit form of 
the velocity distribution function (VDF) as Eq.~(\ref{P_y_fun}), 
which is valid up to numerical accuracy, even for NESSs 
where the LE description breaks down.
The functional form has a velocity-dependent 
``reciprocal temperature'' $\beta_y(v_y)$.
In the limit of large absolute values of $v_y$, 
this reciprocal temperature converges 
to a constant value $\beta_y^{\infty}$, 
which is different from both the equilibrium temperature 
and the kinetic temperature, 
and the VDF thus approaches asymptotically a Maxwell distribution 
with temperature $1/\beta_y^{\infty}$.
We have also found that there exists a crossing value 
of the driving force below (above) which 
$\beta_y^{\infty} > \beta_y(0)$ ($\beta_y^{\infty} < \beta_y(0)$).
It might be difficult to obtain 
such a nontrivial dependence of $\beta_y$ on $E$ 
by a naive perturbation expansion in terms of $E$.

The asymptotic property of the tails of the VDFs in NESSs 
is similar to that of nonequilibrium distribution functions 
of microscopic energy currents in thermal conduction models
\cite{Yukawa_et_al}.
In those models the distribution function 
of energy currents parallel to the temperature gradient 
has a tail for large negative (positive) values of the current 
which asymptotically obeys an equilibrium distribution 
with temperature $T_-$ ($T_+$).
Here, $T_-$ and $T_+$ are different from the local temperature 
at the place where the distribution is measured, 
but equal to the local kinetic temperatures 
at slightly forward and backward regions, respectively.
In the present model of electrical conduction, 
the tails of the VDF asymptotically obey an equilibrium distribution 
with  temperature $1/\beta_y^{\infty}$. 
Although $1/\beta_y^{\infty}$ is equivalent in the tails 
for positive and negative values of velocity 
(because of the isotropy of the model in the direction 
perpendicular to the driving force), 
it differs from both the equilibrium temperature 
and the kinetic temperature.
We therefore expect that it is a universal feature in NESSs 
that the distribution function of a microscopic current 
approaches an equilibrium distribution in the limit of large currents.
This might be related to a representation of the
nonequilibrium distribution of microscopic states, 
which has been derived recently \cite{KNST}.

Unlike the results for thermal conduction models, 
we do not yet know the physical meaning of 
the temperature $1/\beta_y^{\infty}$.
Also we do not understand the meanings of other parameters 
in the functional form of the VDF and the crossing value of the driving force.
These subjects remain for future research.

The author is grateful to A. Shimizu 
for helpful discussions and comments.

\end{document}